\pdfoutput=1
\documentclass[letterpaper,twocolumn,10pt,notitlepage,nofootinbib,superscriptaddress,preprintnumbers]{revtex4-1}

\usepackage{amsfonts}
\usepackage{bm,graphicx}
\usepackage{hyperref}

\pdfoutput=1 
\usepackage{amssymb,amsmath,hyperref}
\usepackage{color}
\usepackage{graphicx}
\usepackage{bm}

\usepackage[utf8]{inputenc}
\usepackage{amsmath,amssymb}
\usepackage{graphicx}
\usepackage{bm}
\usepackage{comment} 
\usepackage{subfigure}
\usepackage{array}
\usepackage{multirow} 
\usepackage{diagbox}
\usepackage{verbatim}

\newcommand{\amu}[0]{a_\mu^{\text{HVP,LO}}}
\newcommand{\da}[0]{\Delta a_\mu}
\newcommand{\tr}{\mathrm{tr}\,}
\newcommand{\ul}[1]{\underline{#1}}

\renewcommand{\Re}[0]{\text{Re}}
\renewcommand{\Im}[0]{\text{Im}}
\newcommand{\xcut}[0]{\tau_{\text{c}}}
\renewcommand{\vec}[1]{\boldsymbol{#1}}

\hypersetup{colorlinks, linkcolor = [rgb]{0,0.0,0.75}, citecolor = [rgb]{0,0.0,0.75}, urlcolor = [rgb]{0,0.0,0.75}}

\begin{document}

\preprint{\vbox{\hbox{CERN-TH-2019-051} }}

\title{Finite-volume effects in $(g-2)^{\text{HVP,LO}}_\mu$}

\author{Maxwell T. Hansen}
\email[e-mail: ]{maxwell.hansen@cern.ch}
\affiliation{Theoretical Physics Department, CERN, 1211 Geneva 23, Switzerland}
\author{Agostino Patella}
\email[e-mail: ]{agostino.patella@physik.hu-berlin.de}
\affiliation{Institut f\"ur Physik und IRIS Adlershof, Humboldt-Universit\"at zu Berlin, Zum Großen Windkanal 6, D-12489 Berlin, Germany}

\date{\today}

\begin{abstract}
An analytic expression is derived for the leading finite-volume effects arising
in lattice QCD calculations of the hadronic-vacuum-polarization contribution to
the muon's magnetic moment $\amu \equiv (g-2)_\mu^{\text{HVP,LO}}/2$. For
calculations in a finite spatial volume with periodicity $L$,   $\amu(L)$ admits
a transseries expansion with exponentially suppressed $L$ scaling. Using a
Hamiltonian approach, we show that the leading finite-volume correction scales
as $\exp[- M_\pi L]$ with a prefactor given by the (infinite-volume) Compton
amplitude of the pion, integrated with the muon-mass-dependent kernel. To give a
complete quantitative expression, we decompose the Compton amplitude into the
space-like pion form factor, $F_\pi(Q^2)$, and a multi-particle piece. We
determine the latter through NLO in chiral perturbation theory and find that it
contributes negligibly and through a universal term that depends only on the
pion decay constant, with all additional low-energy constants dropping out of
the integral.
\end{abstract}
\maketitle

\section{Introduction}

The discrepancy between experimental \cite{Bennett:2004pv,Bennett:2006fi} and
theoretical \cite{Jegerlehner:2009ry,Davier:2017zfy,Keshavarzi:2018mgv} values and the ongoing measurements at
Fermilab \cite{Carey:2009zzb,Grange:2015fou,Flay:2016vuw,Hong:2018kqx} and
J-PARC \cite{Shimomura:2015aza,Sato:2017sdn} have motivated various lattice QCD
(LQCD) collaborations to calculate the hadronic contributions to $(g-2)_\mu$,
which currently dominate the theoretical uncertainty
\cite{Blum:2002ii,Burger:2013jya,Chakraborty:2014mwa,Chakraborty:2015ugp,
Blum:2015you,Blum:2016lnc,Chakraborty:2016mwy,Blum:2016xpd,DellaMorte:2017dyu,
Giusti:2017jof,Borsanyi:2017zdw,Giusti:2018mdh,Asmussen:2018ovy,Giusti:2018vrc,
Blum:2018mom,Davies:2019efs,Gerardin:2019rua}.
The relevant contributions divide into hadronic light-by-light, leading-order
hadronic vacuum polarization (LO HVP) and electromagnetic as well as
strong-isospin corrections to the HVP. As the dominant hadronic contribution,
the LO HVP must be determined with sub-percent uncertainties to reach a total
theory uncertainty competitive with the expected experimental precision
\cite{Jegerlehner:2017lbd}. Depending on the central values of the theoretical
and experimental updates, the improved precision on both sides will provide
powerful constraints on, or else strong evidence for, new-physics beyond the
Standard Model.

As the only known, systematically-improvable approach to non-perturbative QCD,
numerical lattice QCD is a natural tool in the determination of the LO HVP where
a systematic and precise value is of great importance. The most common approach
is to estimate $\amu \equiv (g-2)_\mu^{\text{HVP,LO}}/2$ via the integral 
\cite{Bernecker:2011gh}
\begin{equation}
\label{eq:amu}
\amu(T,L) =  \frac{2 \alpha^2}{ m_\mu^2} \int_0^{T/2} dx_0  \,   \widehat{\mathcal K}(m_\mu x_0) G_{T,L}(x_0) \,,
\end{equation}
where $\alpha \approx 1/137$ is the fine-structure constant, $m_\mu$ the muon mass and
\begin{align}
\label{eq:Gdef}
G_{T,L}(x_0) & \equiv - \frac13 \sum_{k=1}^3 \int_{L^3} \! d^3 \vec x \, \langle j_k(x_0, \vec x) j_k(0) \rangle_{T,L} \,, \\[5pt]
\begin{split}
\widehat{\mathcal K}(t) & \equiv  
t^2 -2 \pi  t + (8 \gamma_E - 2) +\frac{4}{t^2}+8 \log (t)  \\[-3pt]
& \hspace{30pt}  - \frac{8 K_1(2 t)}{t}   -8 \int_0^\infty dv \frac{e^{- t \sqrt{ v^2+4} }}{ (v^2+4 )^{3/2}}  \,.
\end{split} 
\end{align}
Here $ j_\mu(x) = \sum_f q_f \overline{\psi}_f(x) \gamma_\mu \psi_f(x)$ is the
Euclidean-signature vector current and $K_1(z)$ a Bessel function. We have used
notation to emphasize that the calculation is performed in a finite-volume $T
\times L^3$ Euclidean spacetime with periodic geometry.

In Eq.~\eqref{eq:amu}  the finite temporal extent is accommodated by cutting the
integral at $T/2$. We leave a detailed analysis of finite-$T$ effects, arising both from the boundary conditions and the treatment of large $x_0$ in the integral,
to a future work.  In this work we consider only the finite-$L$ effects,
defining $\amu(L) \equiv \lim_{T \to \infty} \amu(T,L) $. We will show that this
quantity has only exponentially-suppressed finite-volume effects, and the
suppression is controlled by the pion mass $M_\pi$.  

Even when $T$ is taken very large, the large-$x_0$ region of the
integral in Eq.~\eqref{eq:amu} cannot be calculated from the measured
two-point function because of the well-known exponential degradation of the signal-to-noise ratio. In practice,
one can calculate the two-point function $G_{T \to \infty,L}(x_0)$ for $x_0 <
\xcut$ from numerical simulations (possibly with a mild extrapolation
to saturate the $T \to \infty$ limit), and then use additional inputs to
reconstruct the $x_0 > \xcut$ region. This yields a
decomposition
\begin{gather}
\label{eq:aLsep}
  a_\mu^{\text{recon}}(L) = a_\mu(L|x_0 < \xcut) + a^{\text{recon}}_\mu(L|x_0 > \xcut) \,,
\end{gather}
where the superscript ``recon'' stands for reconstructed.
The first term is calculated by restricting the
integration domain in Eq.~\eqref{eq:amu} to $0 < x_0 < \xcut$ and
by using the measured two-point function. The second term
is obtained from an analogous formula where the integral is taken over $\xcut < x_0 < \infty$ and the reconstructed two-point function is used. 

We will see
that $a_\mu(L|x_0 < \xcut)$ approaches the infinite-volume limit
exponentially fast. On the other hand, $a^{\text{recon}}_\mu(L|x_0 > \xcut)$
 may approach $L \to \infty$ more slowly, depending on the exact prescription used. As an extreme example, if one estimates $G_{\infty,L}(x_0)$ for $x_0 > \xcut$ by summing over a fixed number of finite-volume states, the resulting contribution to the HVP will have power-law $L$-dependence~\cite{Luscher:1986pf,Lellouch:2000pv}. 
In practice more sophisticated procedures are employed and the resulting scaling must be considered on a case by case basis.\footnote{As
explained in Ref.~\cite{Bernecker:2011gh},
 one can use the
Lellouch-L\"uscher formalism~\cite{Luscher:1986pf,Lellouch:2000pv,Meyer:2011um},
or else some model~\cite{DellaMorte:2017dyu}, to extract the time-like pion form
factor in infinite volume, and use this as an input in the spectral
representation to calculate the contribution of states below the four-pion
threshold to $a_\mu(L|x_0 > \xcut)$, directly in infinite volume. In this case one trades the
finite-volume effects for other systematics that depend on the particular chosen
procedure.

}

Our main result, the formula for the leading $\exp[- M_\pi L]$
finite-volume effect to $\amu(L)$, is presented in Sec.~\ref{sec:res}, and is
derived in Sec. \ref{sec:der} by means of a hamiltonian formalism in which
quantization along a spatial direction is used to pick out the complete
functional form non-perturbatively. In Sec.~\ref{sec:imp} we discuss the
implications of our expression for ongoing calculations. We find that the
dominant contribution enters through the space-like pion form factor, and, since
the latter is readily calculated on the lattice, this provides a viable method
for correcting the leading $L$-dependence. We estimate also the dominant
contribution to the finite-volume effects of $a_\mu(L|x_0 < \xcut)$,
which can be useful information when devising a strategy along the lines of
Eq.~\eqref{eq:aLsep}.

 Our results differ from Refs.~\cite{Aubin:2015rzx,Bijnens:2017esv} in that these
 work to a fixed order in chiral perturbation theory (ChPT) whereas our result is the
 full non-perturbative expression, to leading-order in the large $L$ expansion.
 In this regard it is worth emphasizing that the strict chiral expansion is limited by the fact that, at N$^3$LO, the momentum-space vector correlator receives a $Q^6$ contribution that leads to a divergence in the integral defining $\amu$.
 
\section{Result}

\label{sec:res}

We define
\begin{center}
\begin{equation}
\label{eq:daLdef}
\da(L) \equiv \amu(L) - \lim_{L \to \infty} \amu(L) \,,
\end{equation} 
\end{center}
where, as in the introduction, we ignore the effects of the finite temporal extent. 
These scale as $e^{- M_\pi T}$ and $e^{- M_\pi \sqrt{T^2 + L^2}}$. Therefore in the commonly used setup $T=2L$, the finite-$T$ corrections are sub-leading and should be dropped.
The separation is plausible from the perspective of a generic effective field theory. Volume effects can be encoded via position-space propagators, summed over all periodic images. The propagator's form then leads to exponential decay falling according to the image distance multiplied with the pion mass. The detailed proof of this separation, based on the methods of Ref.~\cite{Luscher:1985dn}, is given in a second longer publication.

In Sec.~\ref{sec:der} we show that the leading finite-$L$ corrections are given by
\begin{widetext}
\begin{equation}
     \da(L)
   =  - \frac{2 \alpha^2}{ m_\mu^2} 
\int \frac{d p_3 }{2 \pi }  \frac{ e^{-L \sqrt{M_\pi^2 + p_3^2}} }{4 \pi L}
   \int_0^\infty dx_0 \, \widehat{\mathcal K}(m_\mu x_0)  
   \int \frac{dk_3}{2\pi} \cos (  x_0 k_3 ) \! \! \sum_{q=0,\pm 1} \Re \, T_q(-k_3^2, - k_3p_3) + \mathcal O(e^{- \sqrt{2} M_\pi L}   ) \,,
   \label{eq:DIs}
\end{equation}
\end{widetext}
where $T_q$ is the Compton amplitude
\begin{multline}
T_q(k^2 , k \cdot p) \equiv \\ i \lim_{\vec p' \to \vec p}    \int d^4 x \, e^{i k x}  \langle       \vec p' , q       \vert   \text{T}     \mathcal J_\rho(x) \mathcal J^\rho(0)          \vert        \vec p , q  \rangle_{\infty}
\label{eq:Tdef}
\,,
\end{multline}
in the forward limit. %
Here $   \vert        \vec p , q \rangle$ is the relativistically-normalized state of a single pion with momentum $\vec{p}$ and charge $q$, $k^2=k_0^2-\vec{k}^2$ and $k \cdot p=k_0p_0-\vec{k}\vec{p}$ are the Minkowski squared norm and scalar product. Following the discussion after Eq.~(\ref{eq:daLdef}), the subleading exponential, $e^{- \sqrt{2} M_\pi L}$, arisies from an image displaced in two of the spatial directions.

$\mathcal J_\mu(x)$ is the Minkowski current. In the Schr\"odinger picture this is related to its Euclidean counterpart via
\begin{gather}
   \mathcal{J}_0(\vec{x}) = j_0(\vec{x}) \ , \qquad \mathcal{J}_k(\vec{x}) = -i j_k(\vec{x}) \,,
\end{gather}
and the corresponding Heisenberg operators are
\begin{gather}
  \label{eq:heisenberg:eucl}
   j_\mu(x_0,\vec{x}) = e^{x_0 H} j_\mu(\vec{x}) e^{-x_0 H} \ , \\
   \label{eq:heisenberg:mink}
   \mathcal{J}_\mu(t,\vec{x}) = e^{i t H} \mathcal{J}_\mu(\vec{x}) e^{-i t H} \ .
\end{gather}

\section{Derivation\label{sec:der}}

Define $G_{L_\rho}(x_0)$ exactly as $G_{T,L}(x_0)$ in Eq.~(\ref{eq:Gdef}) but in a volume in which all four directions may differ, i.e.~with $L_0 \times L_1 \times L_2 \times L_3$. Then introduce $\Delta G_3(x_0|L) \equiv  [1 - \lim_{L_{0,1,2} \to \infty} ] G_{L_\rho}(x_0) $ as the finite-volume residue due to compactification in the $3$ direction only.

To determine $\Delta G_3(x_0|L)$, we study $G_{L_\rho}(x_0)$ with geometry $L_\rho =   ( L_\perp,L_\perp,L_\perp, L )$ and quantize along the 3 direction. Defining 
  $
   \ul{x} = (x_1,x_2,x_0) = (x_\perp,x_0) 
   $,
the Hamiltonian representation of the Euclidean two-point function yields
\begin{multline}
   G_{L_\rho}(x_0)
   = - \frac13
   \int_0^{L_\perp} d^2 x_\perp \int_0^L dx_3 \\ \times
   \frac{\tr [ e^{-(L-x_3)H} j_\mu(\ul{x}) e^{-x_3H} j_\mu(\ul{0}) ]}{\tr e^{-LH}}
   \ ,
   \label{eq:finiteL:Hamiltonian}
\end{multline}
where the Hamiltonian has a discrete finite-volume spectrum of states in $L_\perp^3$ and the trace is taken over this Hilbert space. Here we are using $L_\bot$ to ensure that intermediate expressions are well-defined. This will be sent to infinity at the end of the calculation. For simplicity, in this formula we have assumed periodic boundary conditions for gluons and antiperiodic boundary conditions for fermions in the $3$ direction. To account for the commonly-used periodic boundary contitions for fermions one should introduce $(-1)^F$ in all traces, where $F$ is the fermion number. This does not change the leading exponential contribution, since this is due to single-pion, hence bosonic, states.

Let $|n\rangle$ be a basis of simultaneous eigenstates of the Hamiltonian (eigenvalue $E_n$), the momentum (eigenvalue $\vec{p}_n$), the charge (eigenvalue $q_n$) operators. Inserting a complete set of such states in both the numerator and the denominator then gives
\begin{multline}
  G_{L_\rho}(x_0)
   = - \frac13
   \sum_{n,n'} \frac{e^{-LE_n}}{\sum_{n''} e^{-LE_{n''}}}
   \int_0^{L_\perp} d^2x_\perp
   \int_0^L dx_3
   \\ \times
   e^{-x_3 (E_{n'}-E_n)}
   \langle n | j_\mu(\ul{x}) | n' \rangle \langle n' | j_\mu(\ul{0}) | n \rangle
   \ .
   \label{eq:finiteL:I-10}
\end{multline}

The role of the coordinates $x_0$ and $x_3$ in this analysis is potentially confusing. In our final results $x_0$ plays the role of the time coordinate. This is the coordinate of integration in Eq.~\eqref{eq:amu}, typically parametrizing the longest Euclidean direction. Here, to identify the leading $L$-dependence, it is convenient 
to quantize along the $3$ direction. One must only take care that, in any given expression, all energies and all states are consistently defined with respect to the same quantization direction.

Returning to Eq.~(\ref{eq:finiteL:I-10}), the integral over $x_3$ can be calculated explicitly. To avoid the need of separating $E_{n'}=E_n$ terms from the rest, 
we introduce the following identity, which holds for all values of $E_{n}$, $E_{n'}$ %
\begin{multline}
   e^{-LE_n} \int_0^L dx_3 \ e^{-x_3 (E_{n'}-E_n)}
   = \\
   \lim_{\epsilon \to 0^+} \Re \frac{e^{-L(E_n+i\epsilon)}-e^{-L (E_{n'}-i\epsilon)}}{E_{n'}-E_n-2i\epsilon}
   \,.
\end{multline}
Substituting into Eq.~\eqref{eq:finiteL:I-10} and exchanging $n' \leftrightarrow n$ in certain terms, we obtain
\begin{multline}
    G_{L_\rho}(x_0)
   =
   - \frac{1}{3} \lim_{\epsilon \to 0^+}
   \sum_{n} \frac{e^{-L E_n}}{\sum_{n''} e^{-LE_{n''}}}
   \int_0^{L_\perp} d^2x_\perp \\ \times
   \left\{
   \Re
   \langle n | j_\mu(\ul{x}) \frac{e^{-i L \epsilon}}{H-E_n-2i\epsilon} j_\mu(\ul{0}) | n \rangle
   + (\epsilon \to - \epsilon)
   \right\}
   \ .
\end{multline}

This expectation value can be expressed in terms of the (finite-volume) Minkowskian two-point function via
\begin{multline}
   \Re \langle n | j_\mu(\ul{x}) \frac{e^{-i L \epsilon}}{H-E_n-2i\epsilon} j_\mu(\ul{0}) | n \rangle + (\epsilon \to - \epsilon)
   = \\
   \Re \, i e^{-i L \epsilon} \int_{-\infty}^\infty dt \ e^{-2\epsilon |t|} \langle n | \text{T} \mathcal{J}_\mu(t,\ul{x}) \mathcal{J}^\mu(0) | n \rangle
   \ ,
\end{multline}
which is valid for $\epsilon>0$ and can be easily proven using Eq.~\eqref{eq:heisenberg:mink} and integrating over $t$ explicitly. We stress that this is a mathematical identity and the parameter $t$ has no relation to any of the spacetime coordinates in the system.

The expansion about $L \to \infty$ is now straightforward as the exponentials are manifest and one can identify the relevant contribution. Neglecting terms of order $e^{- 2 M_\pi L}$ we reach %
\begin{multline}
 \Delta G_3(x_0|L)
=  
   - \frac{1}{3} \lim_{\epsilon \to 0^+}
   \sum_{M_\pi \le E_n < 2 M_\pi} e^{-L E_n}
   \int_0^{L_\perp} d^2x_\perp \\ \times
   \Re \, i e^{-i L \epsilon} \int_{-\infty}^\infty dt \ e^{-2\epsilon |t|} \langle n | \text{T} \mathcal{J}_\mu(t,\ul{x}) \mathcal{J}^\mu(0) | n \rangle_\text{c}
    \label{eq:finiteL:DG3}
   \ ,
\end{multline}
where the connected expectation value is defined as $\langle n | \mathcal{O}  | n \rangle_\text{c}
   \equiv
   \langle n | \mathcal{O}  | n \rangle
   - \langle 0 | \mathcal{O}  | 0 \rangle$.
At this point we can take the $L_\perp \to  \infty$ limit. This is done by replacing the sum over the states in the one-particle region with the phase-space integral  
\begin{multline}
  \sum_{M_\pi \le E_n < 2 M_\pi}
  e^{-L E_n}
  | n \rangle \langle n |
\ \ \   \underset{L_\bot \to \infty}{\longrightarrow} %
  \\ %
  \sum_{q=0,\pm 1} \int_{E(\vec{p})<2 M_\pi} \! \frac{d^3 \vec p  }{(2\pi)^3 } \frac{  e^{-L E(\vec{p})}}{ 2 E(\vec{p})}
  | \vec{p},q \rangle \langle \vec{p} , q |
  \ ,
\end{multline}
and by replacing the connected expectation value with the forward limit
\begin{multline}
  i \langle n | \text{T} \mathcal{J}_\mu(t,\ul{x}) \mathcal{J}^\mu(0) | n \rangle_\text{c}
  \to
  \\ %
  \int \frac{d^4 k}{(2\pi)^4} \, e^{-i (k_0 t - k_\perp x_\perp - k_3 x_0)} T_q(k^2 , k \cdot p)
  \ ,
\end{multline}
where the definition \eqref{eq:Tdef} has been used. In the $L_\perp \to  \infty$ limit, the integrals over $t$ and $x_\perp$ are readily calculated, yielding delta functions in $k_0=k_\perp=0$, i.e.
\begin{multline}
 \Delta G_3(x_0|L)
=  
   - \frac{1}{3}
   \sum_{q=0,\pm 1} \int_{E(\vec{p})<2 M_\pi} \! \frac{d^3 \vec p  }{(2\pi)^3 } \frac{  e^{-L E(\vec{p})}}{ 2 E(\vec{p})}
   \\ \times  
   \Re \int \frac{d k_3}{2\pi} e^{i k_3 x_0} T_q(-k_3^2,-k_3p_3)
   \ .
\end{multline}
In the final expression note that any contribution to the integrand that is odd in $k_3, p_3 \to -k_3, -p_3$ must integrate to zero, which justifies the replacement $e^{ik_3x_0} \to \cos(k_3x_0)$. To complete the derivation we note that the restriction $E(\vec{p})<2M_\pi$ can be dropped, as this amounts to an error of the same order as terms that we are neglecting. Finally, the integral over $p_\perp$ can be explicitly calculated. We reach
\begin{multline}
    \Delta G_3(x_0|L)
   = - \frac{1}{3}
   \sum_{q=0,\pm 1} \int \! \frac{d p_3}{2\pi} \frac{e^{-L \sqrt{M_\pi^2+p_3^2}}}{4 \pi L}
\\ \times   \int \frac{dk_3}{2\pi} \cos(k_3x_0) \Re T_q(-k_3^2,-k_3p_3)
   \ .
\end{multline}
Multiplying the result by $3$ to account for the three directions with compactification $L$, we conclude Eq.~\eqref{eq:DIs}.\footnote{This last step assumes a decomposition similar to that allowing us to neglect finite-$T$ and is demonstrated in detail in a subsequent publication.}

We close by commenting on different choices of boundary conditions. If fermions satisfy $e^{i\theta}$-periodic boundary conditions~\cite{Bedaque:2004kc,deDivitiis:2004kq}, i.e. $\psi_f(x+L_\rho \hat{\rho}) = e^{i\theta_\rho^f} \psi_f(x)$, Eq.~\eqref{eq:finiteL:Hamiltonian} should be modified by inserting $(-1)^Fe^{i\sum_f \theta_3^f N_f}$ in all traces, where $N_f$ is the number operator for the flavour $f$. In this case, Eq.~\eqref{eq:DIs} is modified by replacing
\begin{gather}
\label{eq:twist}
   3 \sum_{q=0,\pm 1} \Re T_q
   \to
   \sum_{k=1}^3
   \{
   2 \cos (\theta_k^u - \theta_k^d) \Re T_{\pm 1}
   + \Re T_0
   \}
   \ ,
\end{gather}
where we have used $T_{+1} = T_{-1} \equiv T_{\pm 1}$ as follows from charge-conjugation invariance.

\section{Implications \label{sec:imp}} 

Having derived the leading-order functional form of $\da(L)$, we close by considering the implications for ongoing numerical LQCD calculations.
Here we mainly focus on periodic boundary conditions but comment again on the role of twisting below.  For convenience we define the charge-summed Compton amplitude, $T \equiv \sum_q T_q$.

We begin by rewriting Eq.~(\ref{eq:DIs}) as\footnote{We drop terms of order $e^{- \sqrt{2} M_\pi L}$ throughout this section.}
\begin{align}
     \da(L)
   & = - \frac{2 \alpha^2  m_\mu}{\pi L}
   \int_0^\infty dk_3 \ \hat{g}(  k_3/m_\mu) \mathcal{T}''(k_3^2|L) \,,
   \label{eq:daLaltform}
\end{align}
where we have introduced
\begin{gather}
\label{eq:calTdef}
 \mathcal{T}(k_3^2|L)
    \equiv
   \int_{-\infty}^{\infty} \frac{d p_3}{2\pi} \ e^{-L \sqrt{M_\pi^2+p_3^2}} \, \Re T(-k_3^2, - p_3k_3) \,, \\
 \hat{g}(\omega)  \equiv \omega \int_{\omega^2}^\infty \! dy \    \int_{y}^\infty \! \frac{dx}{x^{3/2}}   \frac{16}{\sqrt{x + 4} \, \big ( \sqrt{x + 4} + \sqrt{x} \big )^4} \,,
\end{gather}
and $\mathcal{T}''$ is the second derivative of $\mathcal{T}$ with respect to $k_3^2$.
 
\begin{table}
\begin{tabular}{c | c | c | c}
\multicolumn{4}{c}{$-10^2 \times \da(L)/\amu$ } \\[5pt]
$M_\pi L$ & \ \  $F_{\pi}(Q^2) = 1$ \ \  & \ \  $F_{\pi}(Q^2) =  \frac{1}{1 + Q^2/M^2}$ \ \ & \ \ $\mathcal T^{\text{reg}}$ \ \ \\ \hline
4.0 &  0.639 &  1.26    &   $0.019$  \\
5.0 &  0.579 &   0.852  & $0.005$    \\
6.0 &  0.348  & 0.461 & $0.001$    \\
7.0 &  0.180  & 0.226  & $\vert \cdots \vert < 10^{-3}$         \\
8.0 &  0.0863 & 0.104 & $\vert \cdots \vert < 10^{-3}$ 
\end{tabular}
\caption{Contribution to $\da(L)$ from the $F_\pi(Q^2)$-term for functional forms as indicated. Here we take $m_\mu/M_\pi=106/137$ and $M/M_\pi = 727/137$. As a reference value we take $\amu=700 \times 10^{-10}$. \label{tab:t1}}
\end{table}

\begin{figure}[t!]
 \vspace{10pt}
\begin{center}
\includegraphics[width=0.4\textwidth]{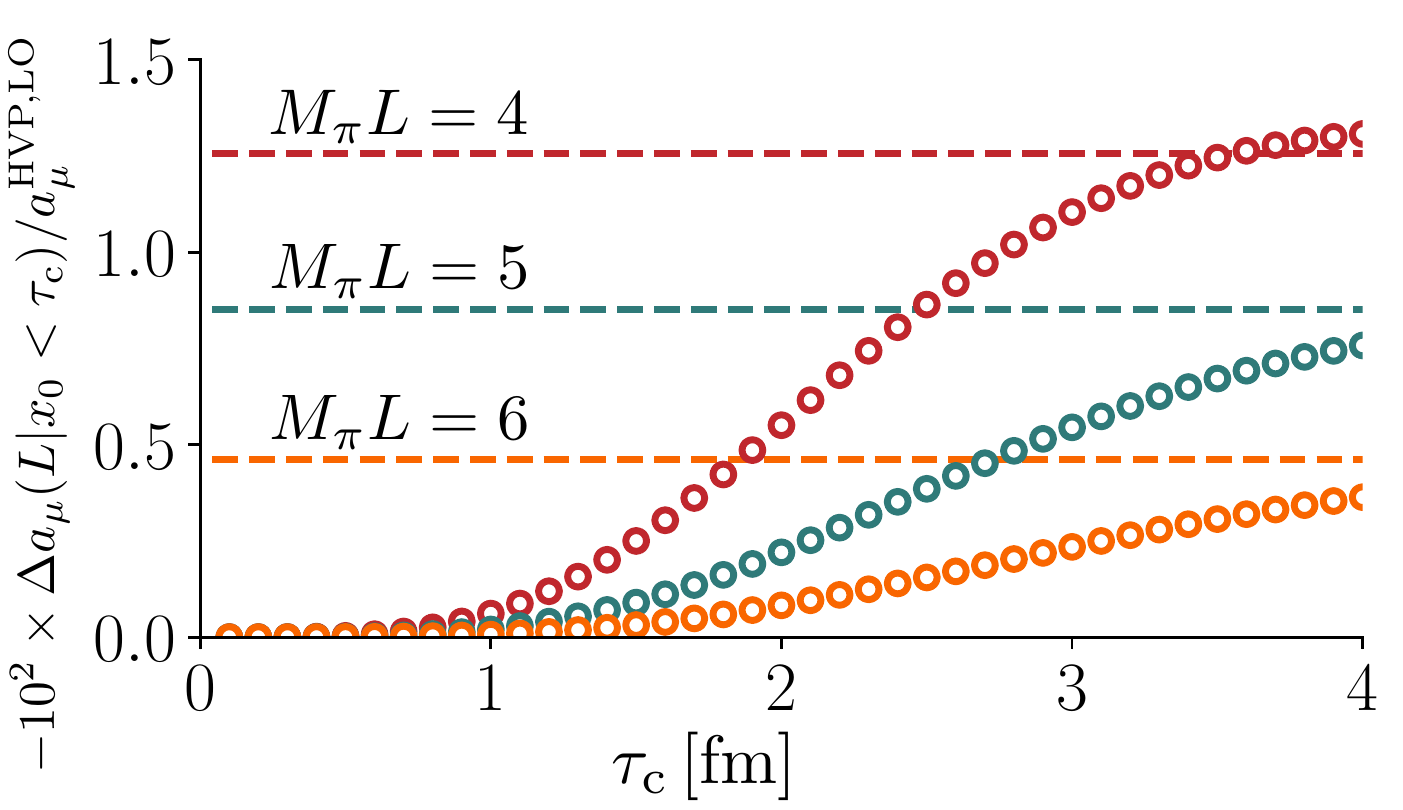}
\caption{Plot of the $F_\pi(Q^2)$-contribution to $\da(L \vert x_0 < \xcut)$ vs.~$\xcut$, taking the monopole ansatz as in Table \ref{tab:t1} and $M_\pi=M_\pi^\text{phys}$, for various values of $M_\pi L$. The horizontal lines give $\xcut = \infty$, i.e.~the full value for $\da(L)$.   \label{fig:cut}}
\end{center}
\end{figure}

We next decompose the Compton amplitude into its pole and analytical contributions
 \begin{multline}
 T(k^2 , k \cdot p)   \equiv T^{\text{reg}}(k^2 , k \cdot p) \\ + \bigg [  \frac{ 2 (4M_\pi^2 -  k^2) F_\pi^2(-k^2)}{- k^2 - 2p \cdot k - i \epsilon}
 +( p \to - p) \bigg ]
 \,,
 \end{multline}
  where $F_\pi $ is the space-like pion form factor and the separation defines $T^{\text{reg}}$. This implies
\begin{equation}
\label{eq:calTsep}
 \mathcal{T}(k_3^2|L) =  2 (4M_\pi^2+k_3^2) F_\pi^2(k_3^2) \zeta(k_3^2|L) +  \mathcal{T}^{\text{reg}}(k_3^2|L) \,,
 \end{equation}
where we have introduced
\begin{equation}
\label{eq:zeta}
 \zeta(k_3^2|L) \equiv    2 \Re \int_{-\infty}^\infty \frac{d p_3}{2\pi} \ \frac{e^{-L \sqrt{M_\pi^2+p_3^2}}}{k_3^2 + 2p_3k_3 - i \epsilon} \,,
\end{equation}
 which can be readily reduced to forms well suited to numerical evaluation.
 The second term in Eq.~(\ref{eq:calTsep}) is given by Eq.~(\ref{eq:calTdef}) with $T \to T^{\text{reg}}$.\footnote{$T^{\text{reg}}(-k_3^2,-p_3k_3)$ is an analytic function in the complex strip defined by $\Im k_3 < M_\pi/2$. In addition, for real $k_3$, both $T^{\text{reg}}(-k_3^2,-p_3k_3)$ and $F(k_3^2)$ are real functions.}  

 In Table \ref{tab:t1} we numerically estimate the contribution of the $F_\pi(Q^2)$-dependent term to $\da(L)$, using two functional forms for the specelike form-factor. We have confirmed that the $F_\pi(Q^2) = 1$ values match the prediction from ChPT. The monopole form, taken from Ref.~\cite{Brommel:2006ww}, %
 is known to describe both experimental and lattice data very well up to $Q^2 = 2.45 \text{GeV}^2$.

We have also calculated the NLO ChPT prediction for $T^{\text{reg}}$ (summed over $\pi_0$ and $\pi^{\pm}$ external states),
  \begin{multline}
 T^{\text{reg}}(- Q^2, k \cdot p) = c_0 + c_1 Q^2  \\ +\frac{    (7 M_\pi^2 + 4Q^2 )   }{6 \pi ^2 f_\pi^2 }  \ z \, \coth^{-1}z  \bigg \vert_{z=  \sqrt{1+4 M_\pi^2/ Q^2 }}  \,,
 \end{multline}
 with the convention that $f_\pi \approx 132 \text{MeV}$. %
 The coefficients $c_0$ and $c_1$ depend on various low-energy constants and on the pion mass. However, the contribution from these terms to $\da(L)$ is identically zero, as can be seen explicitly from Eq.~(\ref{eq:daLaltform}). Evaluating the remaining piece, we find that this contributes negligibly as shown in the third column of Table \ref{tab:t1}. Note that, as can be seen from Eq.~(\ref{eq:twist}), using twisted boundary conditions with $\theta_k^u - \theta_k^d = \pi/2$ sets the $F_\pi(Q^2)$-dependent piece identically to zero, leaving only the contribution from the neutral pion in $T^{\text{reg}}$. 
 This dramatically reduces the leading $L$-dependence.
 
  \begin{table}
\begin{tabular}{c |  c   c   c   c   c}
\multicolumn{6}{c}{$-10^2 \times  \da(L|x_0 < \xcut) /\amu$ } \\[10pt]
\multicolumn{6}{c}{ $M_\pi = M_\pi^{\text{phys}}$} \\[3pt] 
\  $M_\pi L$\   & \    $\xcut =  1\,\text{fm}$  \ & \ \ \ $ 1.5\,\text{fm}$ \ \  \ & \ \ \ $2\,\text{fm}$ \ \ \ & \ \ \ $2.5\,\text{fm}$ \ \  \ &\ \ \ $\infty$ \ \  \ \\ \hline
4.0   & 0.0611   &  0.250 &  0.550 & 0.864  & 1.26 \\
5.0   & 0.0198 & 0.0896 & 0.220 & 0.385 & 0.851 \\
6.0    & 0.00649 & 0.0313 & 0.0825 & 0.155 & 0.461  \\
7.0    & 0.00214 & 0.0108 & 0.0300 & 0.0593 & 0.226 \\
8.0   & 0.00072 & 0.00374 & 0.0108 & 0.0221 & 0.104
\end{tabular}\\[10pt]
\begin{tabular}{c |  c   c   c   c   c}
\multicolumn{6}{c}{ $M_\pi = 2 M_\pi^{\text{phys}}$} \\[3pt] 
\  $M_\pi L$\   & \  $\xcut =  1\,\text{fm}$ \ & \ \ \ $ 1.5\,\text{fm}$ \ \  \ & \ \ \ $2\,\text{fm}$ \ \ \ & \ \ \ $2.5\,\text{fm}$ \ \  \ &\ \ \ $\infty$ \ \  \ \\ \hline
4.0  & 0.231 & 0.682 & 1.08 & 1.28 & 1.38 \\
5.0    & 0.0808 & 0.264 & 0.456 & 0.578 & 0.662   \\
6.0    &   0.0281 & 0.0996 & 0.185 & 0.247 & 0.302 \\
7.0    &  0.00975 & 0.0369 & 0.0727 & 0.102 & 0.134 \\
8.0   & 0.00339 & 0.0135 & 0.0280 & 0.0411 & 0.0576
\end{tabular}\\[10pt]
\begin{tabular}{c |  c   c   c   c   c}
\multicolumn{6}{c}{ $M_\pi = 3 M_\pi^{\text{phys}}$} \\[3pt] 
\  $M_\pi L$\   & \  $\xcut =  1\,\text{fm}$ \ & \ \ \ $ 1.5\,\text{fm}$ \ \  \ & \ \ \ $2\,\text{fm}$ \ \ \ & \ \ \ $2.5\,\text{fm}$ \ \  \ &\ \ \ $\infty$ \ \  \ \\ \hline
4.0  & 0.455 & 1.14 & 1.61 & 1.82 & 1.92 \\
5.0  & 0.162 & 0.430 & 0.634 & 0.730 & 0.778   \\
6.0    & 0.0574 & 0.162 & 0.249 & 0.293 & 0.316 \\
7.0  & 0.0204 & 0.0609 & 0.0970 & 0.117 & 0.128  \\
8.0    & 0.00724 & 0.0227 & 0.0376 & 0.0462 & 0.0515
\end{tabular}
\caption{ Tabulated values of the $F_\pi(Q^2)$-contribution to $\da(L|x_0 < \xcut)$ for various $M_\pi L$, $M_{\pi}$ and $\xcut$. We vary the monople mass according to the result of Ref.~\cite{Brommel:2006ww}: $M^2 = 0.517(23) \text{GeV}^2 + 0.647(30) M_\pi^2$  and hold the reference value fixed at $\amu=700 \times 10^{-10}$. \label{tab:t2}}
\end{table}

When separating the $x_0$ regions as in Eq.~(\ref{eq:aLsep}), it is useful to identify the cut value, $\xcut$, which minimizes the the systematic errors
given by the finite-volume effects of $a_\mu(L|x_0 < \xcut)$, plus
the uncertainties (finite-volume or otherwise) entering through $a^{\text{recon}}_\mu(L|x_0 >
\xcut)$. In Fig.~\ref{fig:cut} we plot the
leading finite-$L$ correction of $a_\mu(L|x_0 < \xcut)$ vs $\xcut$ for various $M_\pi L$.

The same data is presented in Table \ref{tab:t2}, where we additionally vary the pion mass.
At constant $M_\pi L$, increasing the pion mass leads to a decrease in $m_\mu L$
that translates into significantly enhanced volume effects. This behavior is
predicted by an asymptotic expansion in $m_\mu L$ but the latter exhibits poor
convergence so that the dependence is not obvious for these values. Nonetheless
the enhancement is clearly realized in these results, with a contribution
of $\sim 2 \%$ for $\da(L)$ ($\xcut \to \infty$) with $M_\pi
L = 4$ and $M_\pi / M_\pi^{\text{phys}} = 3$.

 \bigskip
 
   \section{Conclusions}

We have presented a fully non-perturbative analysis of the leading finite-$L$
effects in $\amu$. In particular, Eq.~\eqref{eq:DIs} relates the leading
exponential, $\exp[-M_\pi L]$, to the Compton amplitude of an
off-shell photon scattering against a pion in the forward limit. We also argue that the
contribution coming from the one-pion exchange in the Compton
amplitude (corresponding to the two-pion exchange in $\amu$)
is the dominant contribution. We estimate the effect quantitatively using models
for the electromagnetic space-like pion form factor.

The results presented here provide an additional tool for systematically
removing the finite-$L$ effects in $\amu$. One option is to directly improve the
result on each ensemble with a dedicated measurement of $F_\pi(Q^2)$. A
limitation of this analysis is that the neglected $\exp[-
\sqrt{2} M_\pi L]$ terms may not be small. As argued
in~\cite{DellaMorte:2017dyu} this is certainly true in the case of free pions
with $M_\pi L \approx 4$ with leading-exponential domination setting in around $M_\pi L \approx 6$. 
In this vein we also stress that our full, non-perturbative result for the leading exponential can be used to assess and improve predictions, e.g.~from ChPT, by correcting the leading exponential while keeping the fixed-order prediction for the higher exponentials in the series. This is well-motivated since the structure of the pions becomes less important as the exponentials become more suppressed.

On a technical note, it will be interesting to pursue the Hamiltonian method
(already used in~\cite{Lucini:2015hfa}) for identifying finite-$L$ effects in other contexts.

\subsection*{Acknowledgments}
The authors would like to acknowledge and thank Mattia Bruno, Martin L{\"u}scher, Harvey Meyer, and Nazario Tantalo for useful discussions and for helpful comments on a previous version of this manuscript.
\noindent

\bibliographystyle{apsrev} %
\bibliography{refs} %

\begin{thebibliography}{40}
\expandafter\ifx\csname natexlab\endcsname\relax\def\natexlab#1{#1}\fi
\expandafter\ifx\csname bibnamefont\endcsname\relax
  \def\bibnamefont#1{#1}\fi
\expandafter\ifx\csname bibfnamefont\endcsname\relax
  \def\bibfnamefont#1{#1}\fi
\expandafter\ifx\csname citenamefont\endcsname\relax
  \def\citenamefont#1{#1}\fi
\expandafter\ifx\csname url\endcsname\relax
  \def\url#1{\texttt{#1}}\fi
\expandafter\ifx\csname urlprefix\endcsname\relax\def\urlprefix{URL }\fi
\providecommand{\bibinfo}[2]{#2}
\providecommand{\eprint}[2][]{\url{#2}}

\bibitem[{\citenamefont{Bennett et~al.}(2004)}]{Bennett:2004pv}
\bibinfo{author}{\bibfnamefont{G.~W.} \bibnamefont{Bennett}}
  \bibnamefont{et~al.} (\bibinfo{collaboration}{Muon g-2}),
  \bibinfo{journal}{Phys. Rev. Lett.} \textbf{\bibinfo{volume}{92}},
  \bibinfo{pages}{161802} (\bibinfo{year}{2004}), \eprint{hep-ex/0401008}.

\bibitem[{\citenamefont{Bennett et~al.}(2006)}]{Bennett:2006fi}
\bibinfo{author}{\bibfnamefont{G.~W.} \bibnamefont{Bennett}}
  \bibnamefont{et~al.} (\bibinfo{collaboration}{Muon g-2}),
  \bibinfo{journal}{Phys. Rev.} \textbf{\bibinfo{volume}{D73}},
  \bibinfo{pages}{072003} (\bibinfo{year}{2006}), \eprint{hep-ex/0602035}.

\bibitem[{\citenamefont{Jegerlehner and Nyffeler}(2009)}]{Jegerlehner:2009ry}
\bibinfo{author}{\bibfnamefont{F.}~\bibnamefont{Jegerlehner}} \bibnamefont{and}
  \bibinfo{author}{\bibfnamefont{A.}~\bibnamefont{Nyffeler}},
  \bibinfo{journal}{Phys. Rept.} \textbf{\bibinfo{volume}{477}},
  \bibinfo{pages}{1} (\bibinfo{year}{2009}), \eprint{0902.3360}.

\bibitem[{\citenamefont{Davier et~al.}(2017)\citenamefont{Davier, Hoecker,
  Malaescu, and Zhang}}]{Davier:2017zfy}
\bibinfo{author}{\bibfnamefont{M.}~\bibnamefont{Davier}},
  \bibinfo{author}{\bibfnamefont{A.}~\bibnamefont{Hoecker}},
  \bibinfo{author}{\bibfnamefont{B.}~\bibnamefont{Malaescu}}, \bibnamefont{and}
  \bibinfo{author}{\bibfnamefont{Z.}~\bibnamefont{Zhang}},
  \bibinfo{journal}{Eur. Phys. J.} \textbf{\bibinfo{volume}{C77}},
  \bibinfo{pages}{827} (\bibinfo{year}{2017}), \eprint{1706.09436}.

\bibitem[{\citenamefont{Keshavarzi et~al.}(2018)\citenamefont{Keshavarzi,
  Nomura, and Teubner}}]{Keshavarzi:2018mgv}
\bibinfo{author}{\bibfnamefont{A.}~\bibnamefont{Keshavarzi}},
  \bibinfo{author}{\bibfnamefont{D.}~\bibnamefont{Nomura}}, \bibnamefont{and}
  \bibinfo{author}{\bibfnamefont{T.}~\bibnamefont{Teubner}},
  \bibinfo{journal}{Phys. Rev.} \textbf{\bibinfo{volume}{D97}},
  \bibinfo{pages}{114025} (\bibinfo{year}{2018}), \eprint{1802.02995}.

\bibitem[{\citenamefont{Carey et~al.}(2009)}]{Carey:2009zzb}
\bibinfo{author}{\bibfnamefont{R.~M.} \bibnamefont{Carey}} \bibnamefont{et~al.}
  (\bibinfo{year}{2009}).

\bibitem[{\citenamefont{Grange et~al.}(2015)}]{Grange:2015fou}
\bibinfo{author}{\bibfnamefont{J.}~\bibnamefont{Grange}} \bibnamefont{et~al.}
  (\bibinfo{collaboration}{Muon g-2}) (\bibinfo{year}{2015}),
  \eprint{1501.06858}.

\bibitem[{\citenamefont{Flay}(2017)}]{Flay:2016vuw}
\bibinfo{author}{\bibfnamefont{D.}~\bibnamefont{Flay}}
  (\bibinfo{collaboration}{Muon g-2}), \bibinfo{journal}{PoS}
  \textbf{\bibinfo{volume}{ICHEP2016}}, \bibinfo{pages}{1075}
  (\bibinfo{year}{2017}).

\bibitem[{\citenamefont{Hong}(2018)}]{Hong:2018kqx}
\bibinfo{author}{\bibfnamefont{R.}~\bibnamefont{Hong}}
  (\bibinfo{collaboration}{Muon g-2}), in \emph{\bibinfo{booktitle}{{13th
  Conference on the Intersections of Particle and Nuclear Physics (CIPANP 2018)
  Palm Springs, California, USA, May 29-June 3, 2018}}} (\bibinfo{year}{2018}),
  \eprint{1810.03729},
  \urlprefix\url{http://lss.fnal.gov/archive/2018/conf/fermilab-conf-18-551-e.pdf}.

\bibitem[{\citenamefont{Shimomura}(2015)}]{Shimomura:2015aza}
\bibinfo{author}{\bibfnamefont{K.}~\bibnamefont{Shimomura}},
  \bibinfo{journal}{Hyperfine Interact.} \textbf{\bibinfo{volume}{233}},
  \bibinfo{pages}{89} (\bibinfo{year}{2015}).

\bibitem[{\citenamefont{Sato}(2017)}]{Sato:2017sdn}
\bibinfo{author}{\bibfnamefont{Y.}~\bibnamefont{Sato}}
  (\bibinfo{collaboration}{E34}), \bibinfo{journal}{PoS}
  \textbf{\bibinfo{volume}{KMI2017}}, \bibinfo{pages}{006}
  (\bibinfo{year}{2017}).

\bibitem[{\citenamefont{Blum}(2003)}]{Blum:2002ii}
\bibinfo{author}{\bibfnamefont{T.}~\bibnamefont{Blum}}, \bibinfo{journal}{Phys.
  Rev. Lett.} \textbf{\bibinfo{volume}{91}}, \bibinfo{pages}{052001}
  (\bibinfo{year}{2003}), \eprint{hep-lat/0212018}.

\bibitem[{\citenamefont{Burger et~al.}(2014)\citenamefont{Burger, Feng, Hotzel,
  Jansen, Petschlies, and Renner}}]{Burger:2013jya}
\bibinfo{author}{\bibfnamefont{F.}~\bibnamefont{Burger}},
  \bibinfo{author}{\bibfnamefont{X.}~\bibnamefont{Feng}},
  \bibinfo{author}{\bibfnamefont{G.}~\bibnamefont{Hotzel}},
  \bibinfo{author}{\bibfnamefont{K.}~\bibnamefont{Jansen}},
  \bibinfo{author}{\bibfnamefont{M.}~\bibnamefont{Petschlies}},
  \bibnamefont{and} \bibinfo{author}{\bibfnamefont{D.~B.} \bibnamefont{Renner}}
  (\bibinfo{collaboration}{ETM}), \bibinfo{journal}{JHEP}
  \textbf{\bibinfo{volume}{02}}, \bibinfo{pages}{099} (\bibinfo{year}{2014}),
  \eprint{1308.4327}.

\bibitem[{\citenamefont{Chakraborty et~al.}(2014)\citenamefont{Chakraborty,
  Davies, Donald, Dowdall, Koponen, Lepage, and Teubner}}]{Chakraborty:2014mwa}
\bibinfo{author}{\bibfnamefont{B.}~\bibnamefont{Chakraborty}},
  \bibinfo{author}{\bibfnamefont{C.~T.~H.} \bibnamefont{Davies}},
  \bibinfo{author}{\bibfnamefont{G.~C.} \bibnamefont{Donald}},
  \bibinfo{author}{\bibfnamefont{R.~J.} \bibnamefont{Dowdall}},
  \bibinfo{author}{\bibfnamefont{J.}~\bibnamefont{Koponen}},
  \bibinfo{author}{\bibfnamefont{G.~P.} \bibnamefont{Lepage}},
  \bibnamefont{and} \bibinfo{author}{\bibfnamefont{T.}~\bibnamefont{Teubner}}
  (\bibinfo{collaboration}{HPQCD}), \bibinfo{journal}{Phys. Rev.}
  \textbf{\bibinfo{volume}{D89}}, \bibinfo{pages}{114501}
  (\bibinfo{year}{2014}), \eprint{1403.1778}.

\bibitem[{\citenamefont{Chakraborty et~al.}(2016)\citenamefont{Chakraborty,
  Davies, Koponen, Lepage, Peardon, and Ryan}}]{Chakraborty:2015ugp}
\bibinfo{author}{\bibfnamefont{B.}~\bibnamefont{Chakraborty}},
  \bibinfo{author}{\bibfnamefont{C.~T.~H.} \bibnamefont{Davies}},
  \bibinfo{author}{\bibfnamefont{J.}~\bibnamefont{Koponen}},
  \bibinfo{author}{\bibfnamefont{G.~P.} \bibnamefont{Lepage}},
  \bibinfo{author}{\bibfnamefont{M.~J.} \bibnamefont{Peardon}},
  \bibnamefont{and} \bibinfo{author}{\bibfnamefont{S.~M.} \bibnamefont{Ryan}},
  \bibinfo{journal}{Phys. Rev.} \textbf{\bibinfo{volume}{D93}},
  \bibinfo{pages}{074509} (\bibinfo{year}{2016}), \eprint{1512.03270}.

\bibitem[{\citenamefont{Blum et~al.}(2016{\natexlab{a}})\citenamefont{Blum,
  Boyle, Izubuchi, Jin, J{\"u}ttner, Lehner, Maltman, Marinkovic, Portelli, and
  Spraggs}}]{Blum:2015you}
\bibinfo{author}{\bibfnamefont{T.}~\bibnamefont{Blum}},
  \bibinfo{author}{\bibfnamefont{P.~A.} \bibnamefont{Boyle}},
  \bibinfo{author}{\bibfnamefont{T.}~\bibnamefont{Izubuchi}},
  \bibinfo{author}{\bibfnamefont{L.}~\bibnamefont{Jin}},
  \bibinfo{author}{\bibfnamefont{A.}~\bibnamefont{J{\"u}ttner}},
  \bibinfo{author}{\bibfnamefont{C.}~\bibnamefont{Lehner}},
  \bibinfo{author}{\bibfnamefont{K.}~\bibnamefont{Maltman}},
  \bibinfo{author}{\bibfnamefont{M.}~\bibnamefont{Marinkovic}},
  \bibinfo{author}{\bibfnamefont{A.}~\bibnamefont{Portelli}}, \bibnamefont{and}
  \bibinfo{author}{\bibfnamefont{M.}~\bibnamefont{Spraggs}},
  \bibinfo{journal}{Phys. Rev. Lett.} \textbf{\bibinfo{volume}{116}},
  \bibinfo{pages}{232002} (\bibinfo{year}{2016}{\natexlab{a}}),
  \eprint{1512.09054}.

\bibitem[{\citenamefont{Blum et~al.}(2017)\citenamefont{Blum, Christ, Hayakawa,
  Izubuchi, Jin, Jung, and Lehner}}]{Blum:2016lnc}
\bibinfo{author}{\bibfnamefont{T.}~\bibnamefont{Blum}},
  \bibinfo{author}{\bibfnamefont{N.}~\bibnamefont{Christ}},
  \bibinfo{author}{\bibfnamefont{M.}~\bibnamefont{Hayakawa}},
  \bibinfo{author}{\bibfnamefont{T.}~\bibnamefont{Izubuchi}},
  \bibinfo{author}{\bibfnamefont{L.}~\bibnamefont{Jin}},
  \bibinfo{author}{\bibfnamefont{C.}~\bibnamefont{Jung}}, \bibnamefont{and}
  \bibinfo{author}{\bibfnamefont{C.}~\bibnamefont{Lehner}},
  \bibinfo{journal}{Phys. Rev. Lett.} \textbf{\bibinfo{volume}{118}},
  \bibinfo{pages}{022005} (\bibinfo{year}{2017}), \eprint{1610.04603}.

\bibitem[{\citenamefont{Chakraborty et~al.}(2017)\citenamefont{Chakraborty,
  Davies, de~Oliviera, Koponen, Lepage, and Van~de
  Water}}]{Chakraborty:2016mwy}
\bibinfo{author}{\bibfnamefont{B.}~\bibnamefont{Chakraborty}},
  \bibinfo{author}{\bibfnamefont{C.~T.~H.} \bibnamefont{Davies}},
  \bibinfo{author}{\bibfnamefont{P.~G.} \bibnamefont{de~Oliviera}},
  \bibinfo{author}{\bibfnamefont{J.}~\bibnamefont{Koponen}},
  \bibinfo{author}{\bibfnamefont{G.~P.} \bibnamefont{Lepage}},
  \bibnamefont{and} \bibinfo{author}{\bibfnamefont{R.~S.} \bibnamefont{Van~de
  Water}}, \bibinfo{journal}{Phys. Rev.} \textbf{\bibinfo{volume}{D96}},
  \bibinfo{pages}{034516} (\bibinfo{year}{2017}), \eprint{1601.03071}.

\bibitem[{\citenamefont{Blum et~al.}(2016{\natexlab{b}})}]{Blum:2016xpd}
\bibinfo{author}{\bibfnamefont{T.}~\bibnamefont{Blum}} \bibnamefont{et~al.}
  (\bibinfo{collaboration}{RBC/UKQCD}), \bibinfo{journal}{JHEP}
  \textbf{\bibinfo{volume}{04}}, \bibinfo{pages}{063}
  (\bibinfo{year}{2016}{\natexlab{b}}), \bibinfo{note}{[Erratum:
  JHEP05,034(2017)]}, \eprint{1602.01767}.

\bibitem[{\citenamefont{Della~Morte et~al.}(2017)\citenamefont{Della~Morte,
  Francis, G{\"u}lpers, Herdo{\'i}za, von Hippel, Horch, J{\"a}ger, Meyer,
  Nyffeler, and Wittig}}]{DellaMorte:2017dyu}
\bibinfo{author}{\bibfnamefont{M.}~\bibnamefont{Della~Morte}},
  \bibinfo{author}{\bibfnamefont{A.}~\bibnamefont{Francis}},
  \bibinfo{author}{\bibfnamefont{V.}~\bibnamefont{G{\"u}lpers}},
  \bibinfo{author}{\bibfnamefont{G.}~\bibnamefont{Herdo{\'i}za}},
  \bibinfo{author}{\bibfnamefont{G.}~\bibnamefont{von Hippel}},
  \bibinfo{author}{\bibfnamefont{H.}~\bibnamefont{Horch}},
  \bibinfo{author}{\bibfnamefont{B.}~\bibnamefont{J{\"a}ger}},
  \bibinfo{author}{\bibfnamefont{H.~B.} \bibnamefont{Meyer}},
  \bibinfo{author}{\bibfnamefont{A.}~\bibnamefont{Nyffeler}}, \bibnamefont{and}
  \bibinfo{author}{\bibfnamefont{H.}~\bibnamefont{Wittig}},
  \bibinfo{journal}{JHEP} \textbf{\bibinfo{volume}{10}}, \bibinfo{pages}{020}
  (\bibinfo{year}{2017}), \eprint{1705.01775}.

\bibitem[{\citenamefont{Giusti et~al.}(2017)\citenamefont{Giusti, Lubicz,
  Martinelli, Sanfilippo, and Simula}}]{Giusti:2017jof}
\bibinfo{author}{\bibfnamefont{D.}~\bibnamefont{Giusti}},
  \bibinfo{author}{\bibfnamefont{V.}~\bibnamefont{Lubicz}},
  \bibinfo{author}{\bibfnamefont{G.}~\bibnamefont{Martinelli}},
  \bibinfo{author}{\bibfnamefont{F.}~\bibnamefont{Sanfilippo}},
  \bibnamefont{and} \bibinfo{author}{\bibfnamefont{S.}~\bibnamefont{Simula}},
  \bibinfo{journal}{JHEP} \textbf{\bibinfo{volume}{10}}, \bibinfo{pages}{157}
  (\bibinfo{year}{2017}), \eprint{1707.03019}.

\bibitem[{\citenamefont{Borsanyi et~al.}(2018)}]{Borsanyi:2017zdw}
\bibinfo{author}{\bibfnamefont{S.}~\bibnamefont{Borsanyi}} \bibnamefont{et~al.}
  (\bibinfo{collaboration}{Budapest-Marseille-Wuppertal}),
  \bibinfo{journal}{Phys. Rev. Lett.} \textbf{\bibinfo{volume}{121}},
  \bibinfo{pages}{022002} (\bibinfo{year}{2018}), \eprint{1711.04980}.

\bibitem[{\citenamefont{Giusti et~al.}(2018{\natexlab{a}})\citenamefont{Giusti,
  Sanfilippo, and Simula}}]{Giusti:2018mdh}
\bibinfo{author}{\bibfnamefont{D.}~\bibnamefont{Giusti}},
  \bibinfo{author}{\bibfnamefont{F.}~\bibnamefont{Sanfilippo}},
  \bibnamefont{and} \bibinfo{author}{\bibfnamefont{S.}~\bibnamefont{Simula}},
  \bibinfo{journal}{Phys. Rev.} \textbf{\bibinfo{volume}{D98}},
  \bibinfo{pages}{114504} (\bibinfo{year}{2018}{\natexlab{a}}),
  \eprint{1808.00887}.

\bibitem[{\citenamefont{Asmussen et~al.}(2018)\citenamefont{Asmussen, Gerardin,
  Green, Gryniuk, von Hippel, Meyer, Nyffeler, Pascalutsa, and
  Wittig}}]{Asmussen:2018ovy}
\bibinfo{author}{\bibfnamefont{N.}~\bibnamefont{Asmussen}},
  \bibinfo{author}{\bibfnamefont{A.}~\bibnamefont{Gerardin}},
  \bibinfo{author}{\bibfnamefont{J.}~\bibnamefont{Green}},
  \bibinfo{author}{\bibfnamefont{O.}~\bibnamefont{Gryniuk}},
  \bibinfo{author}{\bibfnamefont{G.}~\bibnamefont{von Hippel}},
  \bibinfo{author}{\bibfnamefont{H.~B.} \bibnamefont{Meyer}},
  \bibinfo{author}{\bibfnamefont{A.}~\bibnamefont{Nyffeler}},
  \bibinfo{author}{\bibfnamefont{V.}~\bibnamefont{Pascalutsa}},
  \bibnamefont{and} \bibinfo{author}{\bibfnamefont{H.}~\bibnamefont{Wittig}},
  \bibinfo{journal}{EPJ Web Conf.} \textbf{\bibinfo{volume}{179}},
  \bibinfo{pages}{01017} (\bibinfo{year}{2018}), \eprint{1801.04238}.

\bibitem[{\citenamefont{Giusti et~al.}(2018{\natexlab{b}})\citenamefont{Giusti,
  Lubicz, Martinelli, Sanfilippo, Simula, and Tarantino}}]{Giusti:2018vrc}
\bibinfo{author}{\bibfnamefont{D.}~\bibnamefont{Giusti}},
  \bibinfo{author}{\bibfnamefont{V.}~\bibnamefont{Lubicz}},
  \bibinfo{author}{\bibfnamefont{G.}~\bibnamefont{Martinelli}},
  \bibinfo{author}{\bibfnamefont{F.}~\bibnamefont{Sanfilippo}},
  \bibinfo{author}{\bibfnamefont{S.}~\bibnamefont{Simula}}, \bibnamefont{and}
  \bibinfo{author}{\bibfnamefont{C.}~\bibnamefont{Tarantino}}, in
  \emph{\bibinfo{booktitle}{{36th International Symposium on Lattice Field
  Theory (Lattice 2018) East Lansing, MI, United States, July 22-28, 2018}}}
  (\bibinfo{year}{2018}{\natexlab{b}}), \eprint{1810.05880}.

\bibitem[{\citenamefont{Blum et~al.}(2018)\citenamefont{Blum, Boyle,
  G{\"u}lpers, Izubuchi, Jin, Jung, J{\"u}ttner, Lehner, Portelli, and
  Tsang}}]{Blum:2018mom}
\bibinfo{author}{\bibfnamefont{T.}~\bibnamefont{Blum}},
  \bibinfo{author}{\bibfnamefont{P.~A.} \bibnamefont{Boyle}},
  \bibinfo{author}{\bibfnamefont{V.}~\bibnamefont{G{\"u}lpers}},
  \bibinfo{author}{\bibfnamefont{T.}~\bibnamefont{Izubuchi}},
  \bibinfo{author}{\bibfnamefont{L.}~\bibnamefont{Jin}},
  \bibinfo{author}{\bibfnamefont{C.}~\bibnamefont{Jung}},
  \bibinfo{author}{\bibfnamefont{A.}~\bibnamefont{J{\"u}ttner}},
  \bibinfo{author}{\bibfnamefont{C.}~\bibnamefont{Lehner}},
  \bibinfo{author}{\bibfnamefont{A.}~\bibnamefont{Portelli}}, \bibnamefont{and}
  \bibinfo{author}{\bibfnamefont{J.~T.} \bibnamefont{Tsang}}
  (\bibinfo{collaboration}{RBC, UKQCD}), \bibinfo{journal}{Phys. Rev. Lett.}
  \textbf{\bibinfo{volume}{121}}, \bibinfo{pages}{022003}
  (\bibinfo{year}{2018}), \eprint{1801.07224}.

\bibitem[{\citenamefont{Davies et~al.}(2019)}]{Davies:2019efs}
\bibinfo{author}{\bibfnamefont{C.~T.~H.} \bibnamefont{Davies}}
  \bibnamefont{et~al.} (\bibinfo{collaboration}{Fermilab Lattice,
  LATTICE-HPQCD, MILC}) (\bibinfo{year}{2019}), \eprint{1902.04223}.

\bibitem[{\citenamefont{G{\'e}rardin et~al.}(2019)\citenamefont{G{\'e}rardin,
  C{\`e}, von Hippel, H{\"o}rz, Meyer, Mohler, Ottnad, Wilhelm, and
  Wittig}}]{Gerardin:2019rua}
\bibinfo{author}{\bibfnamefont{A.}~\bibnamefont{G{\'e}rardin}},
  \bibinfo{author}{\bibfnamefont{M.}~\bibnamefont{C{\`e}}},
  \bibinfo{author}{\bibfnamefont{G.}~\bibnamefont{von Hippel}},
  \bibinfo{author}{\bibfnamefont{B.}~\bibnamefont{H{\"o}rz}},
  \bibinfo{author}{\bibfnamefont{H.~B.} \bibnamefont{Meyer}},
  \bibinfo{author}{\bibfnamefont{D.}~\bibnamefont{Mohler}},
  \bibinfo{author}{\bibfnamefont{K.}~\bibnamefont{Ottnad}},
  \bibinfo{author}{\bibfnamefont{J.}~\bibnamefont{Wilhelm}}, \bibnamefont{and}
  \bibinfo{author}{\bibfnamefont{H.}~\bibnamefont{Wittig}}
  (\bibinfo{year}{2019}), \eprint{1904.03120}.

\bibitem[{\citenamefont{Jegerlehner}(2018)}]{Jegerlehner:2017lbd}
\bibinfo{author}{\bibfnamefont{F.}~\bibnamefont{Jegerlehner}},
  \bibinfo{journal}{EPJ Web Conf.} \textbf{\bibinfo{volume}{166}},
  \bibinfo{pages}{00022} (\bibinfo{year}{2018}), \eprint{1705.00263}.

\bibitem[{\citenamefont{Bernecker and Meyer}(2011)}]{Bernecker:2011gh}
\bibinfo{author}{\bibfnamefont{D.}~\bibnamefont{Bernecker}} \bibnamefont{and}
  \bibinfo{author}{\bibfnamefont{H.~B.} \bibnamefont{Meyer}},
  \bibinfo{journal}{Eur. Phys. J.} \textbf{\bibinfo{volume}{A47}},
  \bibinfo{pages}{148} (\bibinfo{year}{2011}), \eprint{1107.4388}.

\bibitem[{\citenamefont{L{\"u}scher}(1986{\natexlab{a}})}]{Luscher:1986pf}
\bibinfo{author}{\bibfnamefont{M.}~\bibnamefont{L{\"u}scher}},
  \bibinfo{journal}{Commun. Math. Phys.} \textbf{\bibinfo{volume}{105}},
  \bibinfo{pages}{153} (\bibinfo{year}{1986}{\natexlab{a}}).

\bibitem[{\citenamefont{Lellouch and L{\"u}scher}(2001)}]{Lellouch:2000pv}
\bibinfo{author}{\bibfnamefont{L.}~\bibnamefont{Lellouch}} \bibnamefont{and}
  \bibinfo{author}{\bibfnamefont{M.}~\bibnamefont{L{\"u}scher}},
  \bibinfo{journal}{Commun. Math. Phys.} \textbf{\bibinfo{volume}{219}},
  \bibinfo{pages}{31} (\bibinfo{year}{2001}), \eprint{hep-lat/0003023}.

\bibitem[{\citenamefont{Meyer}(2011)}]{Meyer:2011um}
\bibinfo{author}{\bibfnamefont{H.~B.} \bibnamefont{Meyer}},
  \bibinfo{journal}{Phys. Rev. Lett.} \textbf{\bibinfo{volume}{107}},
  \bibinfo{pages}{072002} (\bibinfo{year}{2011}), \eprint{1105.1892}.

\bibitem[{\citenamefont{Aubin et~al.}(2016)\citenamefont{Aubin, Blum, Chau,
  Golterman, Peris, and Tu}}]{Aubin:2015rzx}
\bibinfo{author}{\bibfnamefont{C.}~\bibnamefont{Aubin}},
  \bibinfo{author}{\bibfnamefont{T.}~\bibnamefont{Blum}},
  \bibinfo{author}{\bibfnamefont{P.}~\bibnamefont{Chau}},
  \bibinfo{author}{\bibfnamefont{M.}~\bibnamefont{Golterman}},
  \bibinfo{author}{\bibfnamefont{S.}~\bibnamefont{Peris}}, \bibnamefont{and}
  \bibinfo{author}{\bibfnamefont{C.}~\bibnamefont{Tu}}, \bibinfo{journal}{Phys.
  Rev.} \textbf{\bibinfo{volume}{D93}}, \bibinfo{pages}{054508}
  (\bibinfo{year}{2016}), \eprint{1512.07555}.

\bibitem[{\citenamefont{Bijnens and Relefors}(2017)}]{Bijnens:2017esv}
\bibinfo{author}{\bibfnamefont{J.}~\bibnamefont{Bijnens}} \bibnamefont{and}
  \bibinfo{author}{\bibfnamefont{J.}~\bibnamefont{Relefors}},
  \bibinfo{journal}{JHEP} \textbf{\bibinfo{volume}{12}}, \bibinfo{pages}{114}
  (\bibinfo{year}{2017}), \eprint{1710.04479}.

\bibitem[{\citenamefont{L{\"u}scher}(1986{\natexlab{b}})}]{Luscher:1985dn}
\bibinfo{author}{\bibfnamefont{M.}~\bibnamefont{L{\"u}scher}},
  \bibinfo{journal}{Commun. Math. Phys.} \textbf{\bibinfo{volume}{104}},
  \bibinfo{pages}{177} (\bibinfo{year}{1986}{\natexlab{b}}).

\bibitem[{\citenamefont{Bedaque}(2004)}]{Bedaque:2004kc}
\bibinfo{author}{\bibfnamefont{P.~F.} \bibnamefont{Bedaque}},
  \bibinfo{journal}{Phys. Lett.} \textbf{\bibinfo{volume}{B593}},
  \bibinfo{pages}{82} (\bibinfo{year}{2004}), \eprint{nucl-th/0402051}.

\bibitem[{\citenamefont{de~Divitiis et~al.}(2004)\citenamefont{de~Divitiis,
  Petronzio, and Tantalo}}]{deDivitiis:2004kq}
\bibinfo{author}{\bibfnamefont{G.~M.} \bibnamefont{de~Divitiis}},
  \bibinfo{author}{\bibfnamefont{R.}~\bibnamefont{Petronzio}},
  \bibnamefont{and} \bibinfo{author}{\bibfnamefont{N.}~\bibnamefont{Tantalo}},
  \bibinfo{journal}{Phys. Lett.} \textbf{\bibinfo{volume}{B595}},
  \bibinfo{pages}{408} (\bibinfo{year}{2004}), \eprint{hep-lat/0405002}.

\bibitem[{\citenamefont{Brommel et~al.}(2007)}]{Brommel:2006ww}
\bibinfo{author}{\bibfnamefont{D.}~\bibnamefont{Brommel}} \bibnamefont{et~al.}
  (\bibinfo{collaboration}{QCDSF/UKQCD}), \bibinfo{journal}{Eur. Phys. J.}
  \textbf{\bibinfo{volume}{C51}}, \bibinfo{pages}{335} (\bibinfo{year}{2007}),
  \eprint{hep-lat/0608021}.

\bibitem[{\citenamefont{Lucini et~al.}(2016)\citenamefont{Lucini, Patella,
  Ramos, and Tantalo}}]{Lucini:2015hfa}
\bibinfo{author}{\bibfnamefont{B.}~\bibnamefont{Lucini}},
  \bibinfo{author}{\bibfnamefont{A.}~\bibnamefont{Patella}},
  \bibinfo{author}{\bibfnamefont{A.}~\bibnamefont{Ramos}}, \bibnamefont{and}
  \bibinfo{author}{\bibfnamefont{N.}~\bibnamefont{Tantalo}},
  \bibinfo{journal}{JHEP} \textbf{\bibinfo{volume}{02}}, \bibinfo{pages}{076}
  (\bibinfo{year}{2016}), \eprint{1509.01636}.

\end{thebibliography}

\end{document}